\documentclass{optica-article}

\journal{opticajournal} % for journals or Optica Open

\articletype{Research Article}
\newcommand{\ignore}[1]{}
\usepackage{xcolor} % delete this when submitting
\begin{document}

\title{Entanglement Distribution over a Polarization-Stabilized Aerial Fiber}

\author{Yicheng Shi,\authormark{1,*} Jing Su,\authormark{1,2}, Anouar Rahmouni\authormark{1,2}, Pranish Shrestha\authormark{1,3}, Mheni Merzouki\authormark{1}, Gabriel Bello Portmann\authormark{4},Anne Lazenby\authormark{4}, Mael Flament\authormark{4}, Mehdi Namazi\authormark{4}, Abdella Battou\authormark{1}, Oliver Slattery\authormark{1} and Thomas Gerrits\authormark{1}}

\address{\authormark{1}National Institute of Standards and Technology, 100 Bureau Drive, Gaithersburg, Maryland 20899, USA\\
\authormark{2}Joint Quantum Institute, University of Maryland, 3100 Atlantic Building, College Park, Maryland 20742, USA\\
\authormark{3}Department of Computer and Electrical Engineering, Morgan State University, 1700 East Cold Spring Lane, Baltimore, Maryland 21251, USA\\
\authormark{4}Qunnect Inc., 141 Flushing Ave, Suite 1110, Brooklyn, NY 11205, USA}

\email{\authormark{*}yicheng.shi@nist.gov} %% email address is required; see note below about the corresponding author designation

% use {asbstract*} to suppress the copyright line. Copyright information will be added in production

\begin{abstract*} 
Distributing quantum entanglement over telecommunication fiber is a fundamental task in metropolitan-scale quantum networks, enabling advanced applications such as quantum-secured communication, quantum sensing and distributed quantum computing. Although polarization entanglement is relatively easy to generate and manipulate, its distribution across deployed fibers is challenging because of the time-varying polarization transformations imposed by the fiber, which must be actively stabilized. In this work, we experimentally demonstrate the distribution of polarization-entangled photons over a 62-km, partially-aerial fiber between the National Institute of Standards and Technology and the University of Maryland. With polarization stabilization applied to the fiber, we achieved a photon pair rate of approximately $1500$\,$\text{s}^{-1}$ and observed a time-averaged CHSH inequality parameter of $S=2.34\pm0.37$. During a continuous 24-hour experiment, time-multiplexed polarization compensation sessions only cost 7.2\% of the total link operation time, leaving an entanglement distribution uptime of 92.8\%. Our results demonstrate	 the feasibility of distributing polarization-entangled photons over challenging fiber conditions, which is an important step towards the practical deployment of quantum networks.
\end{abstract*}

%%%%%%%%%%%%%%%%%%%%%%%%%%  body  %%%%%%%%%%%%%%%%%%%%%%%%%%
\section{Introduction}
The distribution of entangled photons between spatially separated nodes is a fundamental task in quantum networking, which enables practical applications including quantum-secured communication, quantum sensing and distributed quantum computation~\cite{PhysRevLett.67.661, PhysRevLett.71.1355, PhysRevA.59.4249}. Entanglement distribution links are also the essential building blocks for advanced protocols such as quantum teleportation~\cite{PhysRevLett.70.1895} and entanglement swapping~\cite{PhysRevLett.71.4287}, which ultimately enable quantum repeaters and allow quantum networks to scale to larger sizes~\cite{PhysRevLett.81.5932, RevModPhys.95.045006}. In such networks, entanglement can be realized using different photonic degrees of freedom such as time-bin~\cite{PhysRevA.63.012309}, frequency-bin~\cite{Lu:23}, and polarization~\cite{Ursin2007}. Among these options, polarization entanglement is particularly attractive because it is easy to generate, manipulate, and measure using standard optical components, making it one of the most practical and widely-adopted choices in quantum network development.

%A quantum network interconnects spatially separated nodes to enable the exchange and processing of quantum information, supporting applications in communication, computation, and sensing that exceed classical capabilities. In such a network, photons serve as flying qubits to transmit encoded quantum information as well as to distribute quantum entanglement between network nodes. The latter task, typically referred to as entanglement distribution, is pivotal for quantum networking protocols such as quantum teleportation and entanglement swapping. 

%Entangled states of photons can be created using various photonic degrees of freedom, including phase, time/frequency bin, and polarization. Among these choices, polarization-entangled photons are particularly popular because they are relatively easy to manipulate and measure. The entangled photons can be distributed through either free-space optical links over long distances, or through the existing fiber-optic communication network, which provides excellent connectivity between quantum nodes in a metropolitan area.

However, distribution of polarization-entangled photons over fiber-based quantum networks presents a significant technical challenge. Due to the fiber's routing geometry and birefringence, optical fibers impose random unitary transformations on the photons' state of polarization (SOP) as they propagate~\cite{Rashleigh:78}. These transformations are sensitive to environmental perturbations such as temperature fluctuations and dynamically induced stresses or displacements along the fiber bundle (e.g., from wind or traffic), causing the output SOP to vary over time in an unpredictable manner~\cite{Xavier_2009}. In buried fibers, the polarization transformation can remain stable for several hours or longer~\cite{Wengerowsky2020,Sena:25,10.1063/5.0021755,PRXQuantum.5.030330}, whereas in aerial fibers, the SOP can fluctuate on short time scales of seconds as the fibers are more exposed to the environment~\cite{Li:18,10.1063/5.0225082, Kucera2024}.

In order to maintain stable polarization-entanglement distribution over deployed fiber links, the polarization transformation of the fiber must be continuously monitored and actively stabilized. Several different techniques for polarization stabilization have been developed and can be broadly categorized on the basis of whether or not reference light is used. For fiber links exhibiting slow polarization drift, stabilization is typically achieved using an in-line polarization controller operating in a feedback loop, where the controller is adjusted to minimize a chosen system parameter such as the polarization state visibility or the quantum bit error rate~\cite{Ding:17, Neumann2022, Agnesi:20, Shi:21}. These methods are straightforward to implement and require minimal hardware overhead; however, their achievable stabilization bandwidth is limited by the integration time required to accumulate sufficient measurement statistics for reliable feedback. When the fiber polarization transformation fluctuates on a comparable timescale, these schemes fail to maintain stable polarization transformation across the fiber link.

%but their speed is limited by measurement integration times which may be longer than the timescale at which the polarization may be stable. 

%Such methods are straightforward to implement and require minimal hardware overhead; however, the achievable stabilization bandwidth is constrained by the time required to accumulate sufficient measurement statistics for reliable feedback, which can exceed the characteristic timescale of polarization fluctuations in deployed fibers.

Fast compensation can be achieved by injecting a polarization reference light that co-propagates with the entangled photons through time- or wavelength-division multiplexing. The SOP of the reference light is prepared in two or more non-orthogonal states and undergoes almost the same fiber-induced polarization transformation as the entangled photons. At the fiber output, the transformed SOPs of the reference light are measured and utilized as feedback signals for the polarization controller, which is set to restore the received reference states either through iterative optimization~\cite{Li:18,Kucera2024,Xavier:08,Neumann2022,Chapman:24,Chapman01:24,Agnesi:20,Ding:17,Shi:21} or a single-step inverse transformation~\cite{Wang:09}.

%To the best of our knowledge, the distribution of polarization-entangled photons over aerial fibers, which represents a more realistic scenario for quantum network deployment, has not yet been demonstrated.
To date, most reported entanglement distribution experiments have been carried out in buried optical fibers, which are well isolated and inherently stable~\cite{10.1063/5.0021755,Wengerowsky2020,Sena:25,PRXQuantum.5.030330}. However, a smaller yet non-negligible portion of the existing fiber infrastructure relies on aerial fibers, which experience more challenging environmental conditions. In this work, we demonstrate the distribution of entangled photons over a polarization-stabilized, partially-aerial fiber connecting the National Institute of Standards and Technology (NIST) and the University of Maryland (UMD). This 62\,km link exhibits significantly poorer polarization stability compared with buried fibers and is dynamically stabilized with a pair of active polarization compensation (APC) devices deployed across the link. Over a continuous 24-hour experiment, we measure a time-averaged Clauser-Horne-Shimony-Holt (CHSH) inequality parameter of $S=2.34\pm0.37$ with a violation of the classical bound observed for more than 20\,hours. This experiment demonstrates the feasibility of implementing quantum networks over a challenging, real-world aerial fiber environment.

\section{Experimental setup}
\begin{figure}[htbp]
\centering
\includegraphics[width=\textwidth]{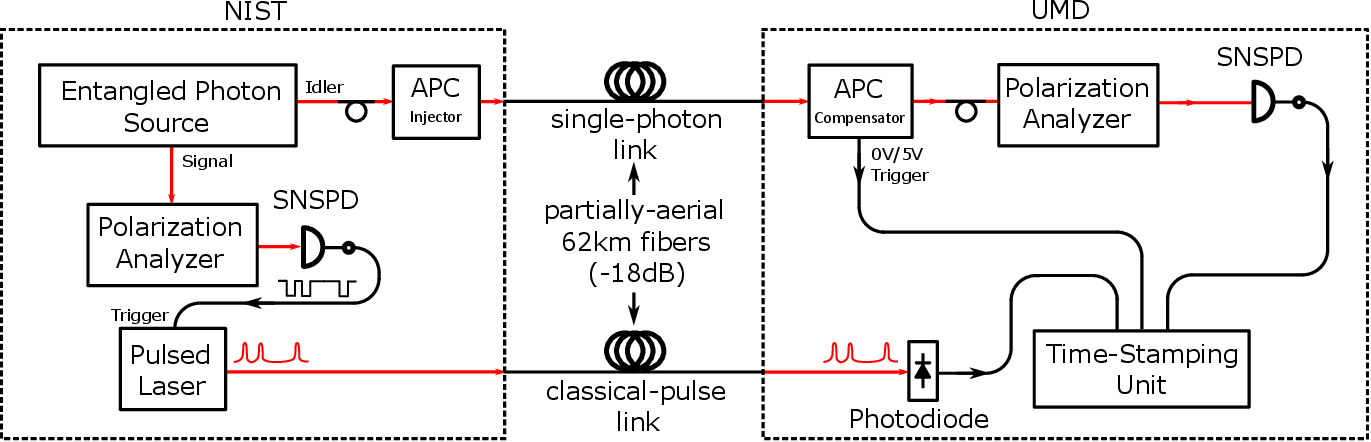}
\caption{Schematic of the entanglement distribution experiment over a 62\,km, partially aerial fiber. Idler photons from the entangled photon pairs are transmitted across the single-photon fiber link, which is polarization-stabilized with a pair of automated polarization compensation modules. The signal photons are detected locally at NIST, and a parallel fiber from the same TX/RX pair transmits the timing information of the signal photons to UMD. The signal-idler coincidence events are recorded with a time-stamping unit located at UMD.}
\label{fig:entanglement_dist_setup}
\end{figure}

A schematic of the entanglement distribution experiment is shown in Fig.~\ref{fig:entanglement_dist_setup}. A polarization-entangled photon source \ignore{(EPS1000, Nucrypt) }located at NIST generates pairs of signal and idler photons initially prepared in the Bell state $|\Phi^+ \rangle$ = $\frac{1}{\sqrt{2}}(|HH\rangle + |VV\rangle)$. The signal and idler photons are spectrally filtered to an optical bandwidth of about 0.3\,nm, centered at 1555.75\,nm and 1549.32\,nm respectively (ITU 100\,GHz grid channels 35 and 27). The signal photons are detected locally at NIST, whereas the idler photons are transmitted to UMD through a 62\,km fiber with a link loss of approximately 18\,dB. The polarization transformation of the fiber link is stabilized by a pair of active polarization compensation modules deployed at NIST and UMD; details of their operation are provided in the following section. A pair of polarization analyzers\ignore{(PA1000, Nucrypt)}, each consisting of variable retarders and polarizing beam splitters, are placed at the two locations to set the polarization measurement basis for the incoming photons.

Superconducting Nanowire Single-Photon Detectors (SNSPDs) are deployed at both sites to detect the incoming photons with $>70\%$ detection efficiencies. At UMD, the detector output is connected to a time-stamping device \ignore{(Time Tagger 20, Swabian Instruments) }to record the arrival times of the idler photons. On the NIST side, the SNSPD photon-detection signals are converted into optical pulses by triggering a 1550\,nm pulsed laser\ignore{(LDH-1550, PicoQuant)}, which are transmitted to UMD through a parallel optical fiber from the same TX/RX pair. At UMD, the received optical pulses are converted back into electrical signals by a high-bandwidth photodiode\ignore{(DET08CFC, Thorlabs)}, and are recorded by the same time-stamping device. Since both fibers belong to the same TX/RX pair, they experience nearly identical variations in optical path length, which maintains a stable relative delay between the signal and idler photon detection times. This "Electrical-Optical-Electrical" link effectively connects the SNSPD output at NIST directly to UMD, enabling photon coincidence identification between distant nodes using a single time-stamping unit. 

\section{Fiber stabilization}
\begin{figure}[h]
	\centering\includegraphics[width=\textwidth, angle=0]{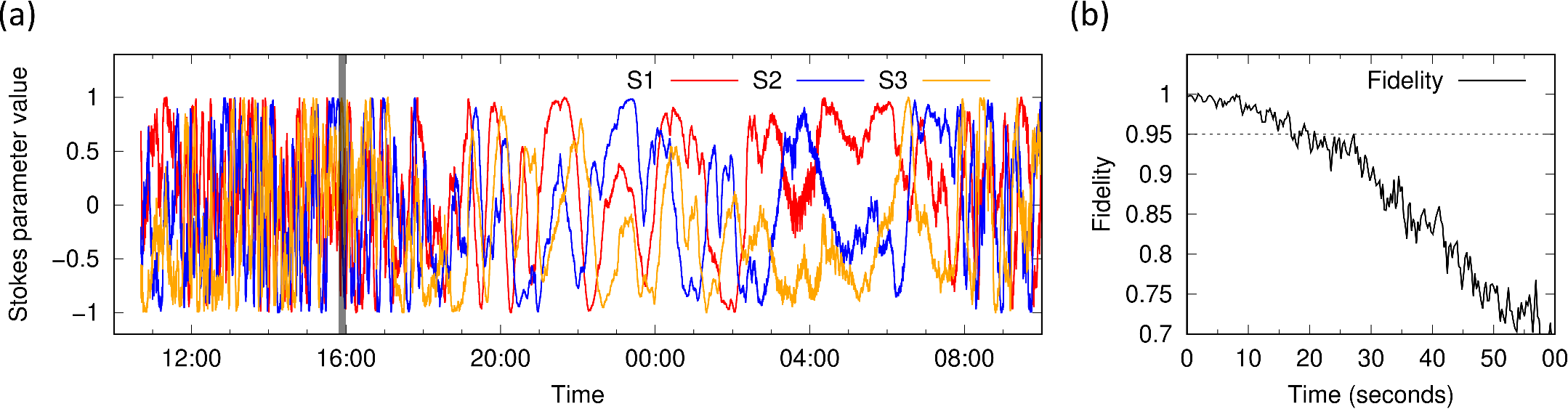}
	\caption{(a) Stokes parameters of probe light measured at the output of the fiber link. (b) Fidelity of the output SOP measured over a 1-minute interval (shaded region in (a)). Fidelity is calculated relative to the SOP at the beginning of the interval ($\text{Fidelity}=1$ at $t=0$).}
	\label{fig:polarization-stability}
\end{figure}

The 62\,km optical fiber connecting NIST and UMD is about 70\% aerial-based and exhibits poor stability compared to buried fibers. To characterize the stability of the link, we injected a polarized probe light and monitored its SOP at the fiber output. The measured Stokes parameters of the transmitted light are shown in Fig.~\ref{fig:polarization-stability}(a), which reveals different rates of polarization fluctuations at different times of the day: at night, the output polarization drifts slowly and can remain stable for over tens of minutes, whereas during daytime the polarization state drifts away by tens of degrees within a matter of seconds. As shown in Fig.~\ref{fig:polarization-stability}(b), the fidelity of the transmitted SOP can drop below 95\% in less than 20\,s during certain periods, rendering the distribution of polarization-entanglement infeasible without active polarization stabilization.

%\subsection*{Active polarization compensation}
\begin{figure}[htbp]
	\centering\includegraphics[width=\textwidth]{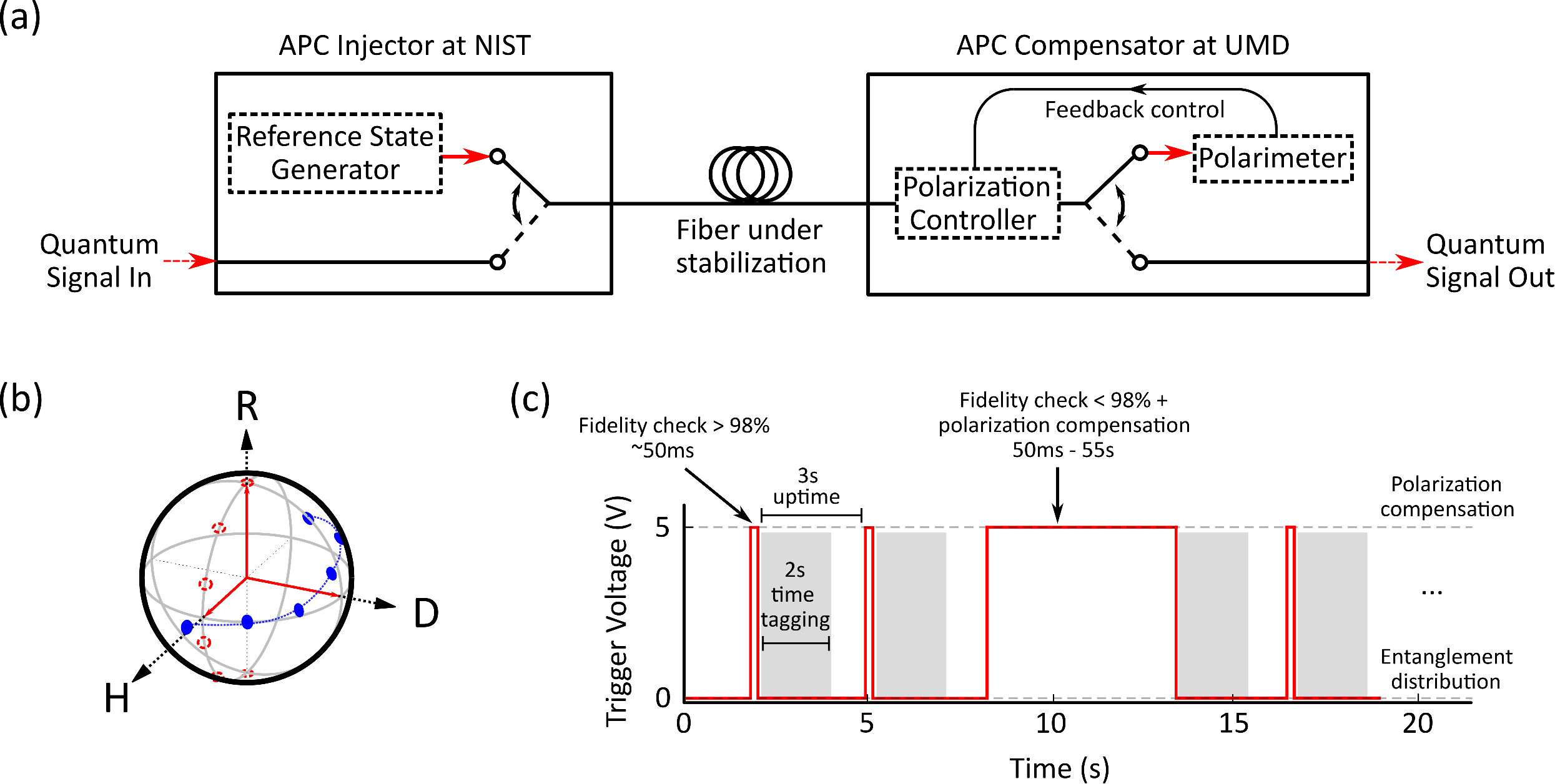}
	\caption{(a) Simplified schematic illustrating the working principle of the APC injector/compensator pair. (b) Sequence of six polarization reference states generated by the APC injector (red) and the corresponding states measured by the APC compensator after fiber transmission (blue). (c) Exemplary timing sequence of the APC compensator trigger voltage level.}
	\label{fig:APC-detail}
\end{figure}
%~\cite{PRXQuantum.5.030330}
The aerial fiber is stabilized by a pair of Automated Polarization Compensation (APC) modules, consisting of an injector and a compensator deployed across the link. A simplified diagram of the APC modules is shown in Fig.~\ref{fig:APC-detail}(a). At NIST, the APC injector launches about 0.5\,mW of laser light into the fiber, whose SOP is cyclically modulated among a sequence of designated reference polarization states, denoted by the red dots in Fig.~\ref{fig:APC-detail}(b).

The polarization-modulated reference signal is generated at the same wavelength as the idler photons (1549.32\,nm) and therefore experiences the same polarization transformation during fiber transmission. At UMD, the transformed reference-signal SOPs, which are denoted by the blue dots in Fig.~\ref{fig:APC-detail}(b), are measured by a polarimeter in the APC compensator. For each set of received reference states, the compensator computes in real time the fidelities relative to the original SOPs, and uses a cost function derived from these fidelities as the error signal to actuate an in-line polarization controller. The feedback loop implements a gradient-descent algorithm to minimize this error signal, aiming to achieve a near-unity fidelity which indicates a fully-compensated fiber channel.

Since the reference signals and the idler photons are generated at the same wavelength, polarization compensation and entanglement distribution must be performed in a time-multiplexed manner. This is achieved using a pair of optical switches in the APC modules, which alternately switch the fiber link between the two operations. The time-multiplexing sequence is illustrated in Fig.~\ref{fig:APC-detail}(c). When polarization compensation begins, the APC injector and compensator take control of the fiber and evaluate the fidelities of the received reference states. If the minimum value of the measured fidelities exceeds our preset threshold of 98\%, the APC takes no action and immediately releases the fiber for entanglement distribution. If this fidelity falls below 98\%, the APC compensator activates the compensation feedback loop and proceeds until the fidelity exceeds a 99\% target threshold, or until a timeout limit of 55\,seconds is reached. After each compensation session, the optical fiber is switched back to the entanglement distribution path for a 3-second measurement period before the next compensation session begins. These 3-second intervals are regarded as the ``uptime'' for the entanglement distribution link, given that the idler photons can only be detected during this time window.

\section{Results}

With an average pump power of 1.2\,mW, we measure approximately $2\times10^5$ photon pairs per second when both signal and idler photons from entanglement source are detected locally using SNSPDs. In the entanglement distribution experiment, the aerial fiber introduces an optical loss of about 18\,dB, while other components---such as fiber connectors and polarization controllers---contribute an additional 3\,dB. As a result, about 1500 pairs of entangled photons are detected between NIST and UMD within a coincidence window of 1.6\,ns.

\begin{figure}[htbp]
\centering\includegraphics[width=\textwidth, angle=0]{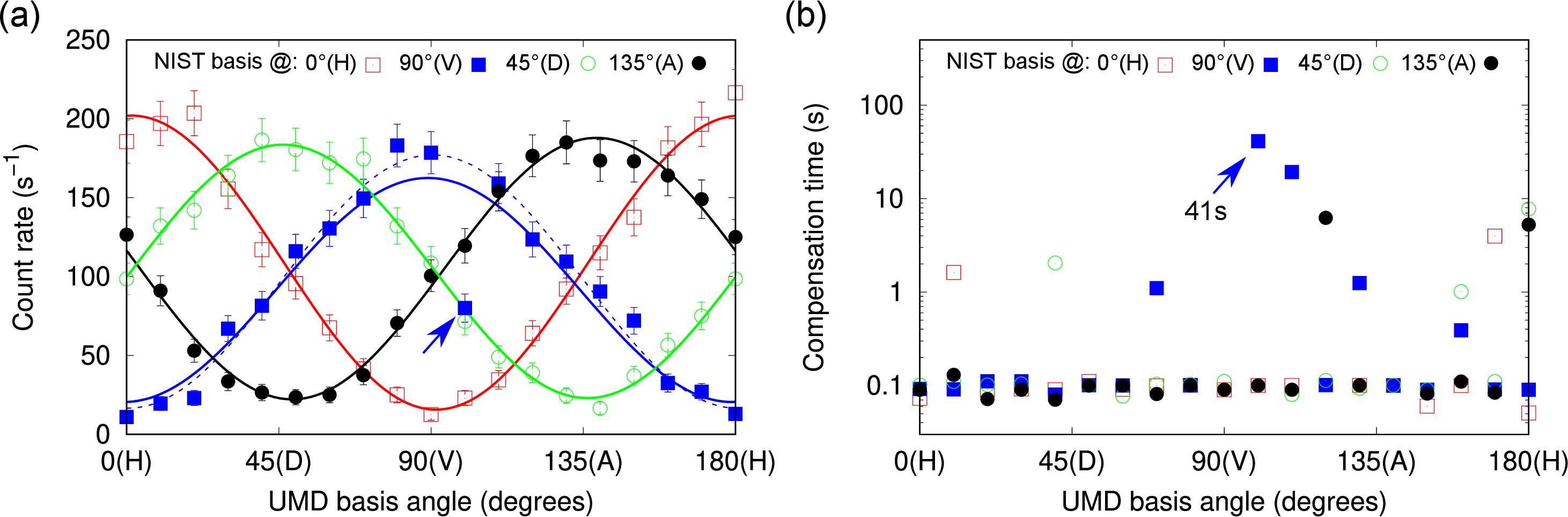}
\caption{(a) Exemplary polarization-correlation fringes measured with active polarization stabilization. For the measurement with NIST analyzer set to 90$^\circ$ (V), the solid blue curve shows the fitted result including all data points, while the dashed blue curve shows the corrected fit obtained by excluding the outlier marked by the blue arrow. (b) Polarization compensation time spent by the APC before recording each data points in (a). The prolonged compensation time (41\,seconds) is likely caused by a temporary, volatile change in the fiber environment.}
\label{fig:fringes}
\end{figure}

\begin{table}[htbp]
	\centering
	\caption{Visibilities and S-value extracted from the fitted curves in Fig.~\ref{fig:fringes}(a).}
	\label{tab:vis}
	\begin{tabularx}{\textwidth}{>{\centering\arraybackslash}X|c|c|c|c}
		\textbf{NIST Analyzer Basis} 
		%\parbox[c][2.5em][c]{3.5cm}{\centering \textbf{NIST Analyzer}\\ \textbf{Basis}}
		& 0$^\circ$ (H) & 45$^\circ$ (D) & 90$^\circ$ (V) & 135$^\circ$ (A) \\
		\hline
		\textbf{Visibility} & 
		$0.87\pm0.02$ & 
		$0.78\pm0.03$ & 
		\parbox[c][2.5em][c]{3.5cm}{\centering $0.76\pm0.03$\\[-0.1em] $0.82\pm0.03$ (corrected)} & 
		$0.79\pm0.03$ \\
		\hline
		\textbf{S-value} & 
		\multicolumn{4}{c}{
			\parbox[c][2.5em][c]{10cm}{\centering $2.26\pm0.04$\\[-0.1em] $2.30\pm0.04$ (corrected)}
		}
	\end{tabularx}
\end{table}

To characterize the polarization-entangled state distributed across the aerial fiber link, we performed a series of polarization-correlation fringe measurements. The received signal and idler photons are sent through the two polarization analyzers located at NIST and UMD, and the photon coincidence rates are measured at different analyzer basis settings. The analyzer at NIST is fixed to four polarization bases: 0$^\circ$(H), 45$^\circ$(D), 90$^\circ$(V) and 135$^\circ$(A). For each of the four settings, the UMD analyzer basis is rotated from 0$^\circ$ to 180$^\circ$ in 10$^\circ$ increments. At each polarization basis configuration, the time-stamping unit remains idle until the ongoing or the upcoming polarization compensation session finishes. Immediately after the end of a session, the time-stamping unit is activated to record signal-idler coincidences for 2 seconds within the 3-second uptime window. During all measurements, uncertainties in the measured coincidence rates are estimated assuming Poissonian statistics for the photon-counting events.

The measured coincidence rate varies as a function of the UMD analyzer's basis angle, which allows us to extract the Clauser-Horne-Shimony-Holt (CHSH) inequality parameter S~\cite{PhysRevLett.23.880} from the curve-fitting results. An exemplary set of measured coincidence rates is shown in Fig.~\ref{fig:fringes}(a), while the visibilities of the corresponding fitted curves~\cite{visibilityNote}, together with the extracted S-value, are summarized in Table \ref{tab:vis}. The measured value of $S=2.26\pm0.04$ exceeds the classical bound of $|S|=2$ and signifies the presence of polarization-entanglement between the distributed signal and idler photons during this particular measurement.

Fig.~\ref{fig:fringes}(b) shows the compensation time spent by the APC before recording each data point in Fig.~\ref{fig:fringes}(a). While most compensation sessions are completed within a few hundreds of milliseconds, some longer sessions can last over tens of seconds. These prolonged sessions are likely caused by rapid fluctuations in the fiber environment (e.g. vibrations caused by wind gusts or traffic) that exceed the bandwidth of the APC feedback loop, and may persist even after a session terminates. In such cases, a compensation session may either fail to reach the 99\% target fidelity and time out, or terminate successfully while the fiber transformation continues to drift during the subsequent uptime window, thereby degrading the measurement outcomes. For example, a 41-second compensation session in Fig.~\ref{fig:fringes}(b) corresponds to an outlier in Fig.~\ref{fig:fringes}(a) (indicated by the blue arrow) which results in a reduced visibility of $0.76\pm0.03$. Excluding this data point from curve-fitting increases the corresponding visibility to $0.82\pm0.03$ and improves the extracted $S$-value from $2.26\pm0.04$ to $2.30\pm0.04$.

\begin{figure}[htbp]
	\centering\includegraphics[width=\textwidth, angle=0]{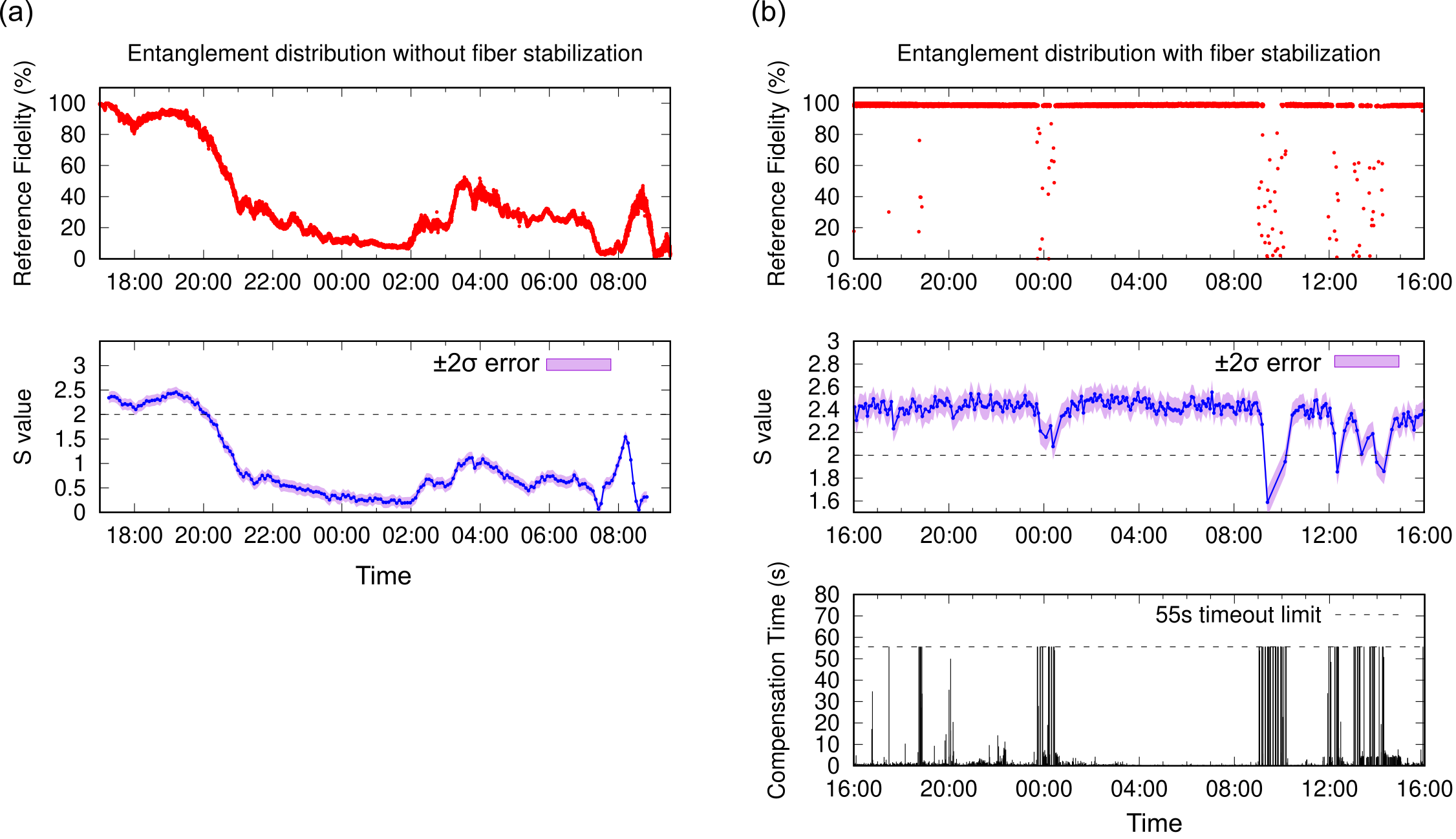}
	\caption{(a) Minimum fidelity of the received APC reference states and $S$-parameter measured over 16\,hours without polarization stabilization. (b) Minimum reference state fidelity, S-parameter value and polarization compensation time recorded over 24\,hours with polarization stabilization enabled.}
	\label{fig:sparam}
\end{figure}

Long-term performance of the polarization-stabilized aerial fiber was characterized by continuously monitoring both the minimum reference-state fidelity and the S-parameter value during entanglement distribution. Fig.~\ref{fig:sparam}(a) shows the measurement results obtained without polarization stabilization. In this case, the reference-state fidelity quickly degraded to below 95\% within a matter of minutes. The S-parameter value dropped below the classical bound of 2 after about three hours. No violation of the CHSH inequality was observed for the remainder of the 16-hour measurement.

In contrast, with polarization stabilization enabled, the reference-state fidelity was
maintained above 98\% for most of the measurement period as shown in Fig.~\ref{fig:sparam}(b). Over the full 24-hour experiment, we measure a time-averaged S-parameter of $2.34\pm0.37$, exceeding the classical bound. Occasional drops in reference-state fidelity and the corresponding S-parameter value are mostly observed during daytime measurements and coincide with the occurrence of prolonged compensation sessions, which indicate environmental disturbances that temporarily exceeded the capability of the APC. By excluding data acquired after these timed-out compensation sessions, the averaged S-parameter improves to $S=2.39\pm0.14$. Despite these disturbances, the APC spent only about 7.2\% of the total operation time to perform polarization stabilization, leaving an effective 92.8\% uptime for entanglement distribution over the stabilized aerial fiber link. 

\section{Conclusion}

We have experimentally demonstrated the distribution of polarization-entangled photons between two distant laboratories connected by a 62-km, partially-aerial fiber link. In this experiment, the time-varying polarization transformation of the fiber was actively stabilized using a pair of automated polarization compensation modules, which were set to maintain a minimum fidelity of 98\% for the time-multiplexed polarization reference states. Over a continuous 24-hour measurement period, we observed a time-averaged CHSH parameter of $S=2.34\pm0.37$, clearly indicating the presence of polarization entanglement between the distributed photons. Despite the unstable fiber condition, polarization compensation required only 7.2\% of the total operation time, leaving a link uptime of 92.8\% for entanglement distribution.

Our experiment demonstrates the feasibility of distributing polarization-entangled photons over challenging fiber conditions, which is an advancement towards the realistic implementation of quantum networks. Although aerial-based optical fibers are typically considered as unsuitable quantum channels for transmitting polarization-encoded or polarization-entangled photons, we show that stable distribution of polarization-entanglement is achievable if the fiber link is actively stabilized.

\begin{backmatter}
%\bmsection{Funding}
%\textcolor{red}{XXX}
\bmsection{Acknowledgment}
The authors thank Cory Nunn and Changhoon Park for insightful discussions and feedback on this manuscript. This work was supported by the Quantum Information Science Initiative (QISI) and National Quantum Initiative (NQI). Contributions of NIST, an agency of the U.S. government, are not subject to copyright.
\bmsection{Disclaimer}
Certain commercial equipment, instruments, or materials are identified in this paper to foster understanding. Such identification does not imply recommendation or endorsement by the National Institute of Standards and Technology, nor does it imply that the materials or equipment identified are necessarily the best available for the purpose.
%\bmsection{Disclosures}
%\textcolor{red}{XXX}
\end{backmatter}

%sample.bib
\bibliography{references}

\begin{thebibliography}{10}
\bibitem{visibilityNote}
The visibility is calculated as $V=(N_{max} - N_{min})/(N_{max} + N_{min})$,
  where $N_{max}$ and $N_{min}$ are the maximum and minimum coincidence rates
  obtained from the fitted curves.
\end{thebibliography}

\end{document}